\documentclass[twocolumn,showpacs,pra]{revtex4}
\usepackage{amssymb}
\usepackage{amsmath}
\usepackage{txfonts}
\usepackage{amsfonts}
\usepackage{graphicx}

\newcommand{\av}[1]{\left\langle{#1}\right\rangle}

\begin{document}

\title{Stochastic resonance in a tristable optomechanical system}
\author{Bixuan Fan and Min Xie}
\affiliation{College of physics, electronics and communication,
Jiangxi Normal University, Nanchang, 330022, China }

\pacs{PACS number}

\begin{abstract}
In this work we theoretically investigate the stochastic resonance (SR) effect in an optomechanical membrane system subject to two weak signals (one optical field and one mechanical force). The quadratic optomechanical coupling allows us to find a region with tristability where the noise activated stochastic switching among three stable states occurs and SR phenomena are observed at the cooperation of input signals and noises. We show that the mechanical force and  the optical field respectively serve as an additive signal and a multiplicative signal to the membrane position and they induce completely different SR behaviors. Moreover, when two signals coexist the SR effect can be enhanced and the beating effect appears in the SR synchronization process with unsynchronized signals.
\end{abstract}

\maketitle
\section{Introduction}
Noises usually play negative roles in signal detection and transfer. However, in some occasions, noises have constructive effects, such as noise induced signal amplification \cite{Wellens1}, noise induced synchronization \cite{Nakao} and noise induced noise suppression \cite{Vilar}. These counterintuitive phenomena occurs mostly in nonlinear systems and they mostly results from the interplay of system nonlinearities and noises. One good example is the phenomenon of stochastic resonance (SR) \cite{Benzi,Gammaitoni}. SR says that noises can improve the output signal quality for a signal passing through a nonlinear system and the optimal improvement is achieved at an optimal noise level. It was first discovered in investigating the periodicity of ice ages in 1981 \cite{Benzi} and since then it been intensively studied in various systems \cite{Gammaitoni,Wellens}, such as physics, chemistry, biology and environmental sciences. SR has been intensively applied to weak signal amplification and detection \cite{Gammaitoni,Wellens}, especially in sensing a faint signal from a noisy background. One well studied model of SR is a classical bistable system with a subthreshold periodic weak signal and noises. Recently, considerable attention has been drawn to SR in monostable \cite{Vilar2,Agudov} and multistable \cite{Arathi,Ghosh,Ghosh2,Saikia} systems. In particular, the study of SR in a triple-well system is important for understanding and manipulation of the escape of a particle from a metastable state \cite{Ghosh,Ghosh2}.

In the past years, cavity optomechanics as a field connecting quantum optics and mechanical objects has attracted increasing research interest \cite{Aspelmeyer}. The radiation pressure mediated optomechanical coupling leads to various nonlinear behaviors, such as bistability \cite{Duan}, multistability \cite{Marquardt}, instability \cite{Arcizet} and chaotic motion \cite{Carmon}. Among a variety of setups of optomechanics, the membrane-in-middle optomechanical system \cite{Thompson,Jayich} is very unique since the mechanical position displacement is quadratically coupled to the optical field unlike the linear relation in most of other optomechanical setups. This nonlinear quadratic coupling allows detecting individual jumps of phonon number in the mechanical mode \cite{Gangat} and adding more interesting physics, for example, mechanical squeezing \cite{Nunnenkamp}, to optomechanical systems. In \cite{Meystre} Meystre group pointed out that in a membrane-in-middle optomechanical system, both double-well and triple-well potential functions of mechanics can be realized by choosing suitable parameters. Thus, this type of system is ideal for studying the SR effect with tristability.

Motivated by these studies, in this work we investigate the SR phenomena in a practical tristable optomechanical membrane system subject to two completely different types of signals (mechanical and optical signals).  Compared to SR in bistable optomechanical systems \cite{monifi,Mueller}, SR in our model displays various periodic switching behaviors. We study the SR effect in three scenarios: a single signal, two synchronized signals, and two unsynchronized signals. We find that a single mechanical signal and a single optical signal induce distinguishing periodic switching behaviors due to their different roles to the membrane position. Besides,  we show that the cooperation of two signals reduces signal amplitudes required for SR to occur and the existence of frequency difference between signals leads to a beating phenomenon in SR, which can be clearly seen from the dynamics and the power spectrum. We know that beating effect usually occurs for two nearly coherent sources, i.e., two lights with small frequency difference. Here the beating occurs between one mechanical force and one optical field with the assistance of the nonlinear optomechanical interaction. We also show that the existing of beating signals largely reduces the sensitivity of the system to initial phase fluctuations of signals at the cost of reduced signal-to-noise ratio (SNR).

The present paper is organized as follows. In Sec. II we introduce our model and provide the steady-state solutions and the stability analysis. In Sec. III, the main signatures of the SR effect including the input-output synchronization and the SNR enhancement are presented. Then, we discuss the interplay of the two signals in the SR process in Sec. IV. Finally, we summarize
our work in Sec. V.

\section{Model}

\begin{figure}[Hbt]
  \centering
  \includegraphics[width=3.0in]{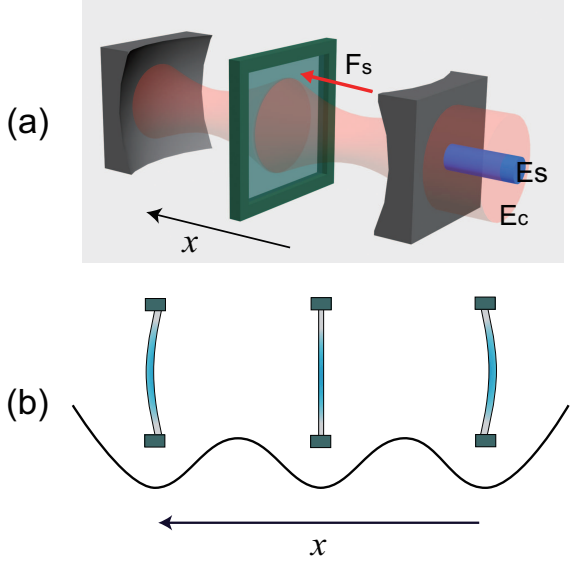}
  \caption{(Color online) A tristable optomechanical membrane system. (a) Model: A thin partially transparent membrane is suspended in the middle of an optical resonator. The photons tunnel through the membrane and the membrane is displaced by the photon induced pressure. The optical resonator is driven by a strong control field $E_c$ and a weak signal field $E_s$ and the membrane is forced by a weak force $F_s$. (b) A cartoon illustration of the triple-well potential function of the steady-state position of the mechanical membrane.}                                                                                              \label{schematic}
  \end{figure}

We consider such a system: a partially transparent dielectric membrane is placed in the middle of a single-mode optical cavity and it divides the cavity into left and right parts as shown in Fig. \ref{schematic}. The membrane position is quadratically coupled to the cavity field. There are three input signals to the system: the optical cavity is driven by a strong control field $E_c$ with frequency $\omega_c$ and a weak signal field $E_s$ with frequency $\omega_s$; a weak force $F_s$ with frequency $\omega_f$ is acted on the membrane. In a rotating frame at the control field frequency $\omega_c$, the Hamiltonian of the described system is given by
\begin{eqnarray}
H &=&\frac{1}{2}\omega _{m}\left( \hat{x}^{2}+\hat{p}^{2}\right) +\Delta \hat{a}^{\dagger }\hat{a}+g\hat{a}^{\dagger }\hat{a}\hat{x}^{2}\\\nonumber
&+&E_c\left(\hat
{a}+ \hat{a}^{\dagger }\right)+E_{s}\left( e^{-i\delta t}\hat{a}^{\dagger }+e^{i\delta t}\hat{a}%
\right)\\\nonumber
&+& F_s \rm{cos}(\omega_f t)\hat{x},
\end{eqnarray}
where $\hat{x}$ and $\hat{p}$ are the dimensionless position and momentum operators of the mechanical mode with frequency $\omega _{m}$; $\hat{a}^{\dagger }\left( \hat{a}\right)$ are creation(annihilation) operators of the optical cavity mode with frequency $\omega _{ca}$; $\Delta = \omega_{ca}-\omega_c$ is the detuning of the optical cavity mode from the control field; $\delta=\omega_s-\omega_c$ is the frequency difference between the optical signal field and control field; $g$ is the quadratic optomechanical coupling strength.

The equations of motion for the expectation values of the system operators $\alpha=\av{\hat{a}}$, $x=\av{\hat{x}}$ and $p =\av{\hat{p}}$ are:
\begin{eqnarray}
\dot{\alpha} &=&-\left( i\Delta +\kappa \right) \alpha -igx^{2}\alpha
-iE_{c}-iE_{s}e^{i\delta t} \\
\dot{\alpha}^{\ast } &=&\left( i\Delta -\kappa \right) \alpha ^{\ast
}+igx^{2}\alpha ^{\ast }+iE_{c}+iE_{s}e^{-i\delta t}\\
\dot{p} &=&-\gamma _{m}p-\omega _{m}x-2g\left\vert \alpha \right\vert
^{2}x-F_s\rm {cos}(\omega_f t)+\xi\\
\dot{x} &=&\omega _{m}p
\end{eqnarray}
where we have phenomenologically introduced the mechanical damping rate $\gamma_m $ and the optical decay rate $2\kappa$.
The mechanical mode is suffered by a temperature-dependent stochastic noise $\xi\left( t\right) $, which obeys the correlation $\left\langle
\xi\left( t\right) \xi\left( t^{\prime }\right) \right\rangle
=2D\left( t-t^{\prime }\right) $ with $D$ being the strength of the thermal noise. Here the noise in the optical field is not considered since it has much less influence on system dynamics than the mechanical noise at a low temperature.

In this work we focus on the dynamics of the mechanical mode and hence we write down the equation of motion merely for mechanics as:
\begin{eqnarray}
\ddot{x}+\gamma_m\dot{x}&=&\omega_m(-\omega_m x-2g|\alpha|^2x-F_s\rm {cos}(\omega_f t)+\xi)\label{as}\\\nonumber
&=& \omega_m(-\omega_m x-2g\frac{|E_c+E_s e^{i\delta t}|^2}{(\Delta+gx^2)^2+\kappa^2}x -F_s\rm {cos}(\omega_f t)+\xi)
\end{eqnarray}
Interestingly, it is seen from Eq. (\ref{as}) that the intensity of optical mode $|\alpha|^2$ plays a role of a multiplicative signal to the mechanical position, which allow us to control the mechanical mode through the optical mode. For instance, one can use the optical signal $E_s$ to tune the photon number of the optical mode in cavity and indirectly to assist the manipulation and detection of the mechanical signal $F_s$.

\begin{figure}[Hbt]
  \centering
  \includegraphics[width=3.3in]{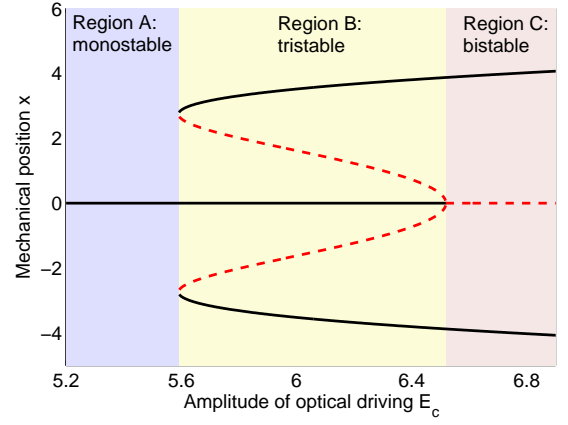}
  \caption{(Color online) System stability diagram. Black solid curves stand for stable branches and red dashed curves for unstable branches. The plane is divided into three regions: monostable region (shadowed by light blue), tristable region (shadowed by light yellow ) and bistable region (shadowed by light red).  The parameters are: $\kappa=5\omega_m$, $\gamma=2.5\omega_m$, $\Delta=3\omega_m$ and $g=-0.4\omega_m$.}                                                                                             \label{stability}
  \end{figure}

We then search the steady-state solutions and study their stability properties in the absence of weak signals $F_s$ and $E_s$. The steady-state solutions can be obtained by setting the time derivatives to zero:
\begin{eqnarray}
0 &=&-\left( i\left( \Delta +gx^{2}_s\right) +\kappa \right) \alpha_s -iE_{c}
\label{S1} \\
0 &=&-\omega _{m}x_s-2g\left\vert \alpha_s \right\vert ^{2}x_s  \label{S2}
\end{eqnarray}%
From Eqs. (\ref{S1})-(\ref{S2}), it is obvious that $x_s=0$ is a solution and the other four possible solutions can be found from the four-order equation of $x_s$:
\begin{eqnarray}
g^2x^4_s+2\Delta gx^2_s+\Delta^2+\kappa^2+2gE_c^2/\omega_m=0
\end{eqnarray}
 $x^{2}_s$ can be solved directly as
\begin{equation}
x^{2}_s=\frac{1}{g^{2}}\left( -\Delta g\pm A\right)   \label{x2}
\end{equation}
with $A=\sqrt{-g^{2}\left( 2gE_{c}^{2}/\omega _{m}+\kappa
^{2}\right) }$. From Eq. (\ref{x2}) one can see that, if $
2gE_{c}^{2}+\omega _{m}\kappa ^{2}<0$. i.e., $g<-\omega _{m}\kappa
^{2}/2E_{c}^{2}$, $x^2_s$ have two real roots. Further, for $-\Delta g>A$, $x_s$ have four real roots:
\begin{equation}
x_s=\pm \frac{1}{\left\vert g\right\vert }\sqrt{-\Delta g\pm A}.
\end{equation}
In such a situation, the tristability is possible to occur.
Otherwise, $x_s$ have two real roots, $x_s=\pm 1\sqrt{-\Delta g+A}/\left\vert
g\right\vert $ and under this condition one may find bisstability.

It is well known that stability properties of a system can be quantitatively studied by the linear stability analysis method. We linearize the system operators around their steady-state values: $\hat{a} (\hat{a}^\dag) \rightarrow \alpha_{s} (\alpha^*_s)+ \hat{a}(\hat{a}^\dag)$, $\hat{x} \rightarrow x_{s} + \hat{x}$ and $\hat{p}\rightarrow p_{s} +\hat{p}$. Ignoring high-order terms, the linearized equation of motion can be written as
\begin{eqnarray}
\dot{y} = J  \hat{y}+\zeta
\end{eqnarray}
where $y=(\hat{a},\hat{a}^\dag,\hat{x},\hat{p})^{T}$, $\zeta=(-iE_se^{i\delta t},iE_se^{-i\delta t},-F_s\rm {cos}(\omega_f t)+\xi,0)^{T}$, and the Jacobian matrix $J$ is given by
\begin{eqnarray}
\centering
J=\left(
                             \begin{array}{cccc}
                               -i(\Delta+gx_{s}^2)-\kappa & 0 & 0 & -i2g\alpha_{s} x_{s} \\
                               0 & i(\Delta+g x_{s}^2)-\kappa & 0 & i2g\alpha^*_{s} x_{s} \\
                               -2g\alpha^*_{s}x_{s} & -2g\alpha_{s}x_{s} & -\gamma_m & -\omega_m-2g|\alpha_{s}|^2 \\
                               0 & 0 & \omega_m & 0 \\
                             \end{array}
                           \right)
\end{eqnarray}
The criterion of a stable solution is that real parts of all eigenvalues of the Jacobian matrix are negative.

 The stability diagram is shown in Fig. 2. The bifurcation curve looks like a deformed pitchfork bifurcation, which usually appear in systems with symmetry.  In a typical pitchfork bifurcation, there are two regions: one stable fixed point; three fixed points (two stable and one unstable). In our model there is one extra region (Region B in Fig. 2) that three stable fixed points (black curves) and two unstable fixed points (red curves) coexist.  In this region, with certain amount of noise the switching among three stable states may occur. The controlling of bistability or tristability can be easily realized by tuning parameters of the system, i.e., when the detuning $\Delta$ varies  from negative to positive or the amplitude of the control field decreases, the system experience the transition from bistability to tristability. In the following sections we focus on SR in the tristable region since SR in a bistable system has been thoroughly studied.

\section{Stochastic resonance: input-output synchronization and SNR enhancement}
In this section we present main results of the SR phenomena in our tristable system from the two characteristic features: the system's periodic response (input-output synchronization) and the resonance behavior of SNR versus the noise strength.

Figure 3 shows single trajectories of the time evolution of the mechanical position under different driving and noise conditions. As shown in Fig. 3(a), the time series of the signals are divided into two stages (labeled by the bright blue dash-dot line): noise off and noise on. We assume that initially the membrane has no displacement in the x axis, i.e., it is located at the middle well. To observe the noise induced system responses, in all simulations we control the signals below thresholds, i.e., the input signal energy is not strong enough for crossing the potential barriers. In the stage of noise absence, the weak force $F_s$ can only drive small-amplitude intrawell oscillations of the membrane and there is no switching from the middle well to adjacent wells. It is interesting to note that there is no dynamics at all when there is only the optical signal $E_s$ acting on the system. This can be understood from Eq. (5): in the absence of noise $\xi$, giving a vanishing initial position $x$, merely the optical mode $|\alpha|^2$ (the third term of Eq. (5)) can not excite the static membrane.
\begin{figure}[Hbt]
  \centering
  \includegraphics[width=3.4in]{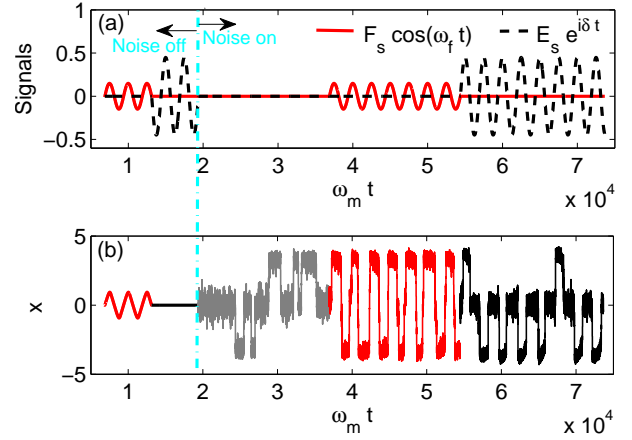}
  \caption{(Color online) Signature of stochastic resonance: the input-output synchronization. (a) Time series of the input signals ($F_s$ and $E_s$). (b) System responses: single trajectories of the steady-state position of the membrane. The dashed-dotted line is used to divide time sequence into two parts: without mechanical thermal noise $D=0$ (left side of the line) and with mechanical thermal noise $D=0.09$ (right side of the line). The detuning $\delta=0.0004\times2\pi\omega_m$ and the mechanical signal frequency $\omega_f=0.0004\times 2\pi\omega_m$, $E_c=5.8\omega_m$, $E_s=0.45\omega_m$ and $F_s=0.15\omega_m$. Other parameters are the same with those in Fig. 2.}                                                                                             \label{SR1}
  \end{figure}
\begin{figure}[Hbt]
  \centering
  \includegraphics[width=3.4in]{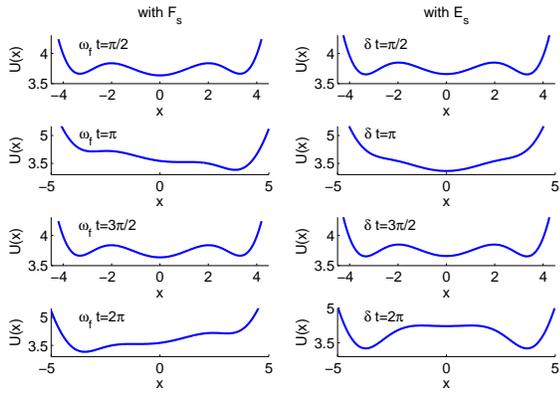}
  \caption{(Color online) Effective potential function of the steady-state position of the membrane at different times of one period. Left panel: the effective potential with the mechanical signal $F_s$; right panel: the effective potential with the optical signal $E_s$. The parameters are the same with Fig. 3.}                                                                                             \label{po1}
  \end{figure}

In the second stage, we add certain amount of mechanical thermal noise ($D=0.09$) to the system. As shown in Fig. 3(b), with the assistance of this noise, the system experiences random switching among three states when there are no input signals $F_s$ and $E_s$. Then, the mechanical signal and optical signal are injected into the system in order. With matched modulation frequencies ($\delta$ and $\omega_f$) and noise strength ($D$), the periodic hopping from middle state to side states is observed and the hopping period follows the input signals. This is a very important signature of SR: the input-output synchronization. It is interesting to note that the periodic hopping behaviours differ for the mechanical and optical signals: the mechanical signal induces symmetric hopping from the middle well to two side wells at the signal frequency $\omega_f$ while the optical signal induces hopping at frequency $\delta$ to side wells but the hopping direction is totally random.

The distinguishing SR behaviours induced by the mechanical signal and the optical signal results from their different relations to the position of the membrane: the mechanical signal $F_s$ is an additive signal directly acting on the membrane while the optical signal $E_s$ indirectly acts on $x$ in the multiplicative way. Their difference can be explained from the effective potential function of the steady-state position of the mechanical resonator U(x), which can be easily derived from Eq.(6) ($\ddot{x}+\gamma_m\dot{x}=-\partial U(x,t)/\partial x$):
\begin{eqnarray}
U\left( x\right) &=&\frac{1}{2}\omega _{m}x^{2}+\frac{|E_{c}
+E_s e^{i\delta t }|^{2}}{\kappa }%
\arctan \frac{\Delta +gx^{2}}{\kappa }\\\nonumber
&+&F_s\rm cos(\omega_f t)x.
\end{eqnarray}
In Fig. \ref{po1}, we show how the effective potential varies during one signal period in the presence of $F_s$ (left panel) and $E_s$ (right panel). Initially the triple-well structure is symmetric with the same depth for each of them. Under the driving of the mechanical signal, the middle well remains unchanged while two side wells oscillate in a synchronized but oppositely-directed manner. Consequently, the steady-state position of the membrane jumps between left and right periodically. In contrast, the optical signal $E_s$ raises and lowers the two barriers simultaneously at the optical signal frequency difference $\delta$. The symmetry of two side wells remains but the depth difference between the middle well and side wells are oscillating. Therefore, the membrane stays at the middle for some time and randomly jumps to one of side wells with the noise activation when the depth of the middle well become small.
\begin{figure}[Hbt]
  \centering
  \includegraphics[width=3.3in]{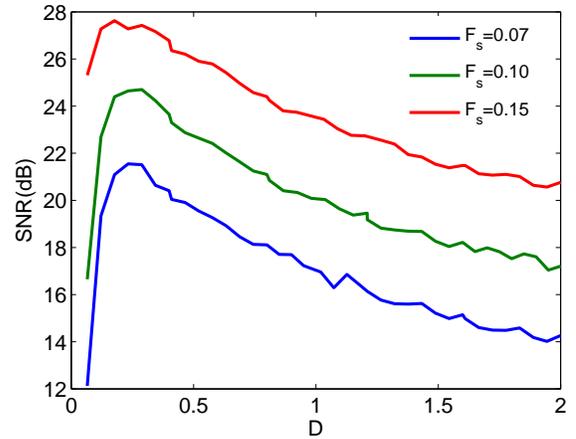}
  \caption{(Color online) Signature of stochastic resonance: SNR versus D for three different amplitudes of the mechanical signal $F_s$. The parameters are the same as those in Fig. \ref{SR1} except for $E_s=0$.}                                                                                             \label{SR2}
  \end{figure}

Another signature of the SR phenomenon--the SNR resonance peak versus noise is presented in Fig. \ref{SR2}. For simplification, here we only consider the situation with the additive signal $F_s$. The SNR is defined as the height of the signal peak in the power spectrum divided by the height of the noise
background at the external driving signal frequency, that is, $\rm{SNR}=P_s/P_n$  \cite{Wellens}.  On the logarithmic scale the SNR is the height of the signal peak above the noise background.

In Fig. \ref{SR2}, we plotted SNR of the output mechanical signal versus the noise strength $D$ for three different signal amplitudes. We can see that all curves have a similar trend: the SNR first rises as the noise increases, reaches a peak, and then decreases as the noise further increases. One may note that there is only a single resonant peak for each SNR curve although the system has multiple stable states, similar to SR in a bistable system \cite{Joshi,Gammaitoni}. This is because that the initial symmetric triple-well potential after acted upon by $F_s$ is still symmetric and hence thermal noise activated switching rates (Kramers rate \cite{Kramers,Hanggi}) to left and right wells are the same, corresponding to a common matched noise level. Besides, the peaks of the SNR curves slightly move towards lower noise side as the signal amplitude $F_s$ increases, which is in good consistence with the discussion in \cite{Gammaitoni}.


\section{Beating phenomena in SR}
So far we have studied system stability properties and SR in the presence of a single mechanical or optical signal. Now we turn to study the situation in the presence of both optical and mechanical signals.

\begin{figure}[Hbt]
  \centering
  \includegraphics[width=3.4in]{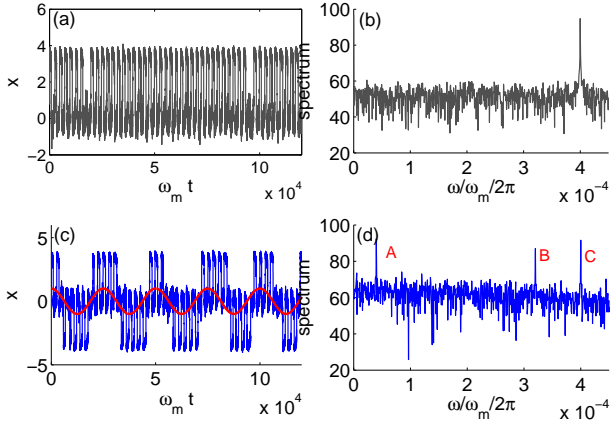}
  \caption{(Color online) The SR synchronizaiton phenomena with two signals $E_s$ and $F_s$ for the situations of (a) $\delta=\omega_f$ and (c)$\omega_f-\delta=\omega_f/10$. (b) and (d) are the Fourier spectrums corresponding to the cases of (a) and (c), respectively. The red curve in (c) is the function of $\rm{cos}((\omega_f-\delta)t)$. The parameters are the same as in Fig. \ref{SR1} except $E_s=0.25$, $F_s=0.08$ and $D=0.05$.}
 \label{beat1}
  \end{figure}

\begin{figure}[Hbt]
  \centering
  \includegraphics[width=3.4in]{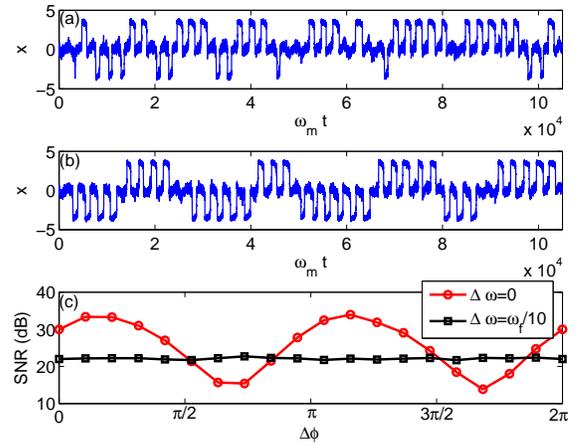}
  \caption{(Color online) Trajectories of the position of the mechanical membrane when the two signals have an initial phase difference $\Delta \phi=\pi/2$ without frequency difference (a)$\Delta\omega=0$ and with (b) $\Delta\omega=\omega_m/10$. (c) represent the SNR as function of $\phi$ for the case with and without frequency difference. The parameters are the same with those in Fig. \ref{beat1}.}
  \label{beat2}
  \end{figure}

When two signals are on, there are two time scales governing the switching dynamics: $\delta$ and $\omega_f$. Note that the time scale of the dynamics induced by $E_s$ is determined by the detuning $\delta$, not its frequency $\omega_s$. From the discussion in last section we know that these two signals periodically modulate the effective potential in two manners with their own modulation frequencies. The modulation would become totally disordered if two modulation frequencies significantly differ from each other. Therefore, to observe nontrivial resonant effect and the synchronization with input signals, the frequency difference $\Delta \omega=\omega_f-\delta$ should not be large. Here we use $\Delta \omega =\omega_f/10$. In Figs. \ref{beat1} and \ref{beat2}, we show how SR phenomena are influenced by the interplay of two signals and how the frequency difference $\Delta\omega$ affects SR.

In Fig. \ref{beat1} (a) we display the SR effect with two synchronized signals $F_s$ and $E_s$ ($\delta=\omega_f=0.0004\times 2\pi\omega_m$).  Here we use two weaker signals $F_s=0.08$ and $E_s=0.25$ and each of them is insufficient to induce the SR effect alone. With the cooperation of two signals, almost deterministic periodic switching from the middle well to the right well is witnessed. This periodicity is much better than that of single signal induced SR in Fig. \ref{S1}. The corresponding spectrum is shown in Fig. \ref{beat1} (b) and a single signal peak is observed at the frequency of $0.0004\times 2\pi\omega_m$, which is exactly the signal frequency. This result is not surprising since we have known that $E_s$ acts as a multiplicative signal and the synchronized additive signal $F_s$ and multiplicative signal $E_s$ would cause this interference of SRs as discussed in Ref. \cite{Ghosh}. In fact, there are many high-order signal peaks in the whole frequency domain due to the optomechanical nonlinearity.

More interesting dynamics is in the situation of a small nonzero modulation frequency difference ($\Delta \omega\neq 0$) as shown in Figs. \ref{beat1}(c) and \ref{beat1}(d). There is a slow envelope along with the fast interwell hopping in the time evolution of $x$. The fast interwell hopping synchronizes with the signals and the slow oscillation actually follows the signal of frequency difference $\rm{cos} (\Delta\omega t)$ (the red sine curve in (c)). This behavior is often referred to as the beating phenomenon. Correspondingly, in the frequency domain (Fig. \ref{beat1}(d)) we can see a main resonance peak at $\omega_f$ and a beating signal peak at $\omega_f-\delta$. One may notice that there is an additional signal at $0.00032\times 2\pi\omega_m$, which is equal to $2\delta-\omega_f $, resulting from the difference-frequency generation process of the second-order of optics and the first-order of mechanics.  Noteworthy, the main SR resonance peak in Fig. \ref{beat1} (d) is lower than that in Fig. \ref{beat1} (b. This is because that the cooperation effect of the signals is strongest when two signals are synchronized. As the frequency difference $\Delta \omega$ approaches 0, the beating signal would gradually vanish and the main SR signal would become stronger. In this sense, the interference result presented in Ref. \cite{Ghosh} is a special case of our model with $\Delta \omega =0$.

We finally study the effect of the initial phase difference $\Delta\phi$ of two signals on the SR phenomena in our symmetric triple-well system. As discussed in Ref. \cite{Ghosh}, for two synchronized input signals ($\Delta \omega=0$), only when their relative initial phase $\Delta\phi=0(\pi)$ the interwell switching is towards one direction and system responses follow the input signals best and For other phases the switching from middle well to two side wells are random due to the splitting of averaged Kramers rates. This means that fluctuations in initial phases may reduce the periodicity of switching and hence reduce the signal quality. To further confirm this effect, we show the time evolution of the membrane position $x$ for $\Delta\phi=\pi/2$ and $\Delta \omega=0$ in Fig. \ref{beat2}(a). It is clear that the hopping from the middle well to the left or right wells is stochastic. However, the situation is dramatically different when the frequency difference of two signals is nonvanishing $\Delta \omega\neq 0$. Fig. \ref{beat2}(b) shows that for the initial phase difference $\Delta\phi=\pi/2$, the beating signal and the synchronization to input signals are still as good as that in Fig. \ref{beat1}(c) with only a slight shift on the time axis. It indicates that the beating effect makes the synchronization behavior of the system robust to the initial phase difference between signals.

To further investigate the influence of the beating effect on the synchronization behavior, we numerically compute the SNR for varying the relative initial phase difference from $0$ to $2\pi$ in Fig. \ref{beat2} (c). For the case of no beating, the SNR curve changes in the cos-like function with an amplitude about 10dB and it is optimized at phase near $n\pi$ ($n=0,1,2$). In contrast, when the frequency beating exists, the SNR almost remains at a stable value with only small fluctuations. The stable value is slightly below the average value of the maximal and minimal of the no beating situation. Thus, the SR effect is more robust to the initial phase fluctuations thanks to the beating phenomenon at the cost of reducing SNR to an average value.

\section{Conclusion}
We investigated SR effects in an optomechanical system with tristability. With the advantage of optomechanics, our model offers more rich nonlinearities and more control channels to study SR. Through the quadratic optomechanical coupling, a weak optical signal can induce SR in the mechanical mode and this SR is significantly different from the mechanical signal induced SR. Besides, we extended the two-synchronized-signal SR model to SR with frequency difference and we observe beating-like phenomena in the SR synchronization. Our results not only present further and deeper understanding of SR in a multistable system, but also introduce one way of controlling mechanical SR phenomena through the optical channel.

Noteworthy, there is more physics about SR to be explored from this model. For instance, at a high operation temperature the noise in the optical field
can not be neglected any more. This noise is indirectly acted on the mechanics in a multiplicative way and one may find new SR behaviors when the optical noise cooperates with the mechanical noise and the two signals.
\begin{acknowledgments}
The authors thanks Dr. Zhenglu Duan  and Dr. Fabing Duan for very helpful discussion. This work is
supported by the National Natural Science Foundation of China under Grants
No. 11504145, No. 11364021, and No.11664014, Natural Science Foundation of Jiangxi
Province under Grants No. 20161BAB211013 and 20161BAB201023.
\end{acknowledgments}


\begin{thebibliography}{99}
\bibitem{Wellens1}Thomas Wellens and Andreas Buchleitner, Phys. Rev. Lett. 84, 5118 (2000).
\bibitem{Nakao} Hiroya Nakao, Kensuke Arai, and Yoji Kawamura, Phys. Rev. Lett. \textbf{98}, 184101 (2007).
\bibitem{Vilar}J. M. G. Vilar and J. M. Rub\'{i}, Phys. Rev. Lett. \textbf{86}, 950 (2001).
\bibitem{Benzi}R. Benzi, A. Sutera, and A. Vulpiani, J. Phys. A \textbf{14}, L453 (1981); R. Benzi, G. Parisi, A. Sutera, and A. Vulpiani, Tellus
\textbf{34}, 10 (1982).
\bibitem{Gammaitoni}Luca Gammaitoni, Peter H\"{a}nggi, Peter Jung, and Fabio Marchesoni, Rev. Mod. Phys., \textbf{70}, 223 (1998).
\bibitem{Wellens} Thomas Wellens, Vyacheslav Shatokhin, and Andreas Buchleitner, Rep. Prog. Phys. \textbf{67}, 45 (2004).
\bibitem{Vilar2}J. M. G. Vilar and J. M. Rub\'{i}, Phys. Rev. Lett. \textbf{77}, 2863 (1996).
\bibitem{Agudov}N. V. Agudov, A. V. Krichigin, D. Valenti, and B. Spagnolo, Phys. Rev. E \textbf{ 81}, 051123 (2010).
\bibitem{Arathi}S. Arathi and S. Rajasekar, Phys. Scr. \textbf{84}, 065011 (2011).
\bibitem{Ghosh}Pulak Kumar Ghosh, Bidhan Chandra Bag, and Deb Shankar Ray, Phys. Rev. E \textbf{75}, 032101 (2007).
\bibitem{Ghosh2} Pulak Kumar Ghosh, Bidhan Chandra Bag, and Deb Shankar Ray, the Journal of Chemical Physics \textbf{127}, 044510 (2007).
\bibitem{Saikia}S. Saikia, A. M. Jayannavar, and Mangal C. Mahato, Phys. Rev. E \textbf{83}, 061121 (2011).
\bibitem{Aspelmeyer}Markus Aspelmeyer, Tobias J. Kippenberg, and Florian Marquardt, Rev. Mod. Phys. \textbf{86}, 1391 (2014).
\bibitem{Duan} Zhenglu Duan, Bixuan Fan, Thomas M. Stace, G. J. Milburn, and Catherine A. Holmes, Phys. Rev. A \textbf{93}, 023802 (2016).
\bibitem{Marquardt} F. Marquardt, J. G. E. Harris, and S. M. Girvin, Phys. Rev. Lett.\textbf{96}, 103901 (2006).
\bibitem{Arcizet}O. Arcizet, P.F. Cohadon, T. Briant, M. Pinard, A. Heidmann, Nature \textbf{444}, 71 (2006).
\bibitem{Carmon}T. Carmon, M. C. Cross, and Kerry J. Vahala, Phys. Rev. Lett.\textbf{98}, 167203 (2007).
\bibitem{Thompson}J. D. Thompson, B. M. Zwickl, A. M. Jayich, Florian Marquardt, S. M. Girvin and J. G. E. Harris, Nature \textbf{452}, 72 (2008).
\bibitem{Jayich}A. M. Jayich, J. C. Sankey, B. M. Zwickl, C. Yang, J. D. Thompson, S. M. Girvin, A. A. Clerk, F. Marquardt and J. G. E. Harris, New Journal of Physics \textbf{10}, 095008 (2008).
\bibitem{Gangat} A. A. Gangat, T. M. Stace, and G. J. Milburn, New J. Phys. \textbf{13}, 043024 (2011).
\bibitem{Nunnenkamp}A. Nunnenkamp, K. Borkje, J. G. E. Harris, and S. M. Girvin, Phys. Rev. A \textbf{82}, 021806(R) (2010).
\bibitem{Meystre}L. F. Buchmann, L. Zhang, A. Chiruvelli, and P. Meystre, Phys. Rev. Lett. \textbf{108}, 210403 (2012).
\bibitem{monifi} Faraz Monifi, Jing Zhang, Sahin Kaya \"{O}zdemir, Bo Peng, Yu-xi Liu, Fang Bo, Franco Nori and Lan Yang, Natu. Phot. \textbf{73}, 1 (2016).
\bibitem{Mueller} F. Mueller, S. Heugel, and L. J. Wan, Phys. Rev. A \textbf{79}, 031804 (2009).
\bibitem{Joshi}Amitabh Joshi and Min Xiao, Phys. Rev. A \textbf{74}, 013817 (2006).
\bibitem{Kramers}H. A. Kramers, Physica \textbf{7}, 284 (1940).
\bibitem{Hanggi}Peter H\"{a}nggi, Peter Talkner, Michal Borkovec, Rev. Mod. Phys. \textbf{62}, 251 (1990).
\end{thebibliography}
\end{document}